\documentclass[a4paper, 11pt]{article}

\usepackage[utf8]{inputenc}
\usepackage{hyperref}
\usepackage{amsmath}
\usepackage{amssymb}
\usepackage{authblk}
\usepackage{lipsum}
\usepackage{geometry}
\usepackage{blkarray}
\usepackage{easybmat}

\usepackage{nicefrac} 
\usepackage{hyperref}
\usepackage{mathtools}

\newcommand{\F}{\mathbb F}

\newtheorem{theorem}{Theorem}[section]
\numberwithin{theorem}{section}

\begin{document}
\pagestyle{empty}

\title{Decoding a class of maximum Hermitian rank metric codes}
\date{}
\author[*]{Wrya K. Kadir}
\author[*]{Chunlei Li}
\author[**]{Ferdinando Zullo\thanks{The research of the last author was supported by the project ``VALERE: VAnviteLli pEr la RicErca" of the University of Campania ``Luigi Vanvitelli'' and by the Italian National Group for Algebraic and Geometric Structures and their Applications (GNSAGA - INdAM).}}
\affil[*]{\small Informatics Department, University of Bergen}
\affil[**]{\small Dipartimento di Matematica e Fisica, Università degli Studi della Campania “Luigi Vanvitelli”}
\maketitle
\begin{abstract}
Maximum Hermitian rank metric codes were introduced by Schmidt in 2018 and in this paper we propose both interpolation-based encoding and decoding algorithms for this family of codes when the length and the minimum distance of the code are both odd. 
\end{abstract}

\thispagestyle{empty}

\section{Introduction}
Let $\mathbb{F}_q^{n\times n}$ be the set of the square matrices of order $n$ defined over $\mathbb{F}_q$, which is the finite field of $q$ elements.
We can equip $\mathbb{F}_q^{n\times n}$ with the following metric
\[ d(A,B)=\mathrm{rk} (A-B), \]
where $\mathrm{rk}(A-B)$ is the rank of the difference matrix $A-B$.
If $\mathcal{C}$ is a subset of $\mathbb{F}_q^{n\times n}$ with the property that for each $A,B \in \mathcal{C}$ then $d(A,B)\geq d$ with $1\leq d\leq n$, then $\mathcal{C}$ is called a \emph{rank metric code with minimum distance} $d$, or that $\mathcal{C}$ is a $d$-\emph{code} \cite{schmidt2018hermitian}.
Furthermore, we say that $\mathcal{C}$ is $\mathbb{F}_q$-\emph{linear} if $\mathcal{C}$ is an $\mathbb{F}_q$-subspace of $(\mathbb{F}_q^{n\times n},+,\cdot)$, where $+$ is the classical matrix addition and $\cdot$ is the scalar multiplication by an element of $\mathbb{F}_q$.

Examples are rank metric codes whose codewords are alternating matrices \cite{delsarte1975alternating}, symmetric matrices \cite{longobardi2020automorphism,schmidt2015symmetric,zhou2020equivalence} and Hermitian matrices \cite{schmidt2018hermitian,trombetti2020maximum}.
In this paper we deal with the latter case.

Consider the conjugation map $\overline{\cdot} $ from $\mathbb{F}_{q^2}$ to itself: $x \mapsto \overline{x} = x^q$.
Let $A\in \mathbb{F}_{q^2}^{n\times n}$ and denote by $A^*$ its conjugate transpose, that is $A^*$ is obtained by the transposition of the matrix $A$ in which the  conjugate map is applied to all of its entries.
A matrix $A\in \mathbb{F}_{q^2}^{n\times n}$ is said to be \emph{Hermitian} if $A^*=A$.
Denote by ${\rm H}_{n}(q^2)$ the set of all Hermitian matrices of order $n$ over $\mathbb{F}_{q^2}$.
In \cite[Theorem 1]{schmidt2018hermitian}, Schmidt proved that if $\mathcal{C}$ is an $\mathbb{F}_q$-linear rank metric code with minimum distance $d$ contained in ${\rm H}_{n}(q^2)$, then
\[|\mathcal{C}| \leq q^{n(n-d+1)}.\]
When the equality is achieved, we say that $\mathcal{C}$ is a \emph{maximum Hermitian $d$-code} or a \emph{maximum Hermitian rank metric code}.

Each rank metric code can be described equivalently in terms of $q$-polynomials.
Let $\mathcal{L}_{n,q}$ denote the quotient $\F_q$-algebra of the algebra of linearized polynomials over $\F_{q^n}$ with respect to $(x-x^{q^n})$, i.e.
\[\mathcal{L}_{n,q}=\left\{ \sum_{i=0}^{n-1} a_i x^{q^i} \colon a_i \in \F_{q^n} \right\}. \]
It is well known that $\mathcal{L}_{n,q}$ is isomorphic to the $\mathbb{F}_q$-algebra of square matrices over $\mathbb{F}_q$.  Using this fact and following \cite{longobardi2020automorphism}, the set ${\rm H}_n(q^2)$ of Hermitian matrices of order $n$ over $\mathbb{F}_{q^{2}}$ can be identified as the following set of linearized polynomials
\[{\mathcal{H}}_n(q^2)=\left\{ \sum_{i=0}^{n-1} c_i x^{q^{2i}} \colon c_{n-i+1}=c_i^{q^{2n-2i+1}} \,\text{ for } \, i \in \{0,\ldots,n-1\} \right\}\subseteq \mathcal{L}_{n,q^2}, \]
where the indices are taken modulo $n$. Note that if $n$ is odd then $c_{(n+1)/2}$ belongs to $\mathbb{F}_{q^n}$.

\smallskip

There are some examples of maximum Hermitian $d$-codes, see \cite{schmidt2018hermitian,trombetti2020maximum}.
The first two examples were given in \cite[Theorems 4 and 5]{schmidt2018hermitian}. 
In this abstract, we consider the decoding of the following class of codes.

\begin{theorem}\cite[Theorem 5]{schmidt2018hermitian}\label{Th-Schmidt-5}
Let $n$ and $d$ be odd integers satisfying $1\leq d\leq n$. 
The Hermitian forms 
$H:\F_{q^{2n}}\times \F_{q^{2n}}\rightarrow \F_{q^2}$ given by $H(x,y)=\mathrm{Tr }\left(y^qL(x)\right)$ with
\begin{equation}\label{ex:E}
  L(x)= (b_0x)^{q^{(n+1)}}+ \sum_{j=1}^{\frac{n-d}{2}} \left( (b_jx)^{q^{(n+2j+1)}}+b_j^{q} x^{q^{(n-2j+1)}} \right),
\end{equation}
as $b_0$ ranges over $\F_{q^n}$ and $b_1,\ldots,b_{\frac{n-d}{2}}$ range over $\F_{q^{2n}}$,
form an additive $d$-code in $\mathrm{H}_n(q^2)$, the set of  $n\times n$ Hermitian matrices over $\F_{q^2}$, and for $z\in \mathbb{F}_{q^{2n}}$ we define $\mathrm{Tr}(z)=z+z^q+\ldots+z^{q^{n-1}}$  which stands for the trace function $\mathrm{Tr}:\F_{q^{2n}}\rightarrow \F_{q^2}$. 
\end{theorem}
The above statement also holds when one extends the $q^2$-polynomials to $q^{2s}$-polynomials with the integer $s$ satisfying $\gcd(s, 2n) = 1$.

Our work in this abstract is also motivated by the recent work of  De La Cruz, Evilla and {\"O}zbudak \cite{de2021hermitian}, where in Section 6 they suggested studying decoding algorithm for Hermitian rank metric codes. A summary of the existing interpolation-based decoding algorithms of rank metric codes is given in \cite[Section V.]{Kadir-li-Zullo21}. 

\section{Encoding and Decoding of Hermitian MRD codes}
\subsection{Encoding}
Assume that $\{\alpha_1,\ldots,\alpha_n\}$ is an $\F_{q^2}$-basis of $\F_{q^{2n}}$ such that $\mathrm{Tr}(\alpha_i^{q} \alpha_j)$ is zero if $i\ne j$ and $1$ if $i=j$.
The property we require on the basis $\{\alpha_1,\ldots,\alpha_n\}$ naturally extends the notion of \emph{self-dual basis} in this setting, for this reason we will call it a \emph{Hermitian self-dual basis} and such a basis always exists, see e.g. \cite[Theorem 4.1]{bose1966hermitian}. Throughout what follows, we denote $x^{[i]}=x^{q^{2i}}$ for the simplicity of presentation.

Let $L(x)$ as in Theorem \ref{Th-Schmidt-5}. For the Hermitian form in Theorem \ref{Th-Schmidt-5},  we have 
\[ 
H(x,y)=\mbox{Tr}(y^q L(x))=\mbox{Tr}(x^q L(y)).
\]
Let $x,y \in \F_{q^{2n}}$, then $x=\sum_{i=1}^n x_i \alpha_i$ and $y=\sum_{i=1}^n y_i \alpha_i$, for some $x_i,y_i \in \F_{q^2}$. It is clear that $\mathrm{Tr}(x^qy)=\langle(x_1^q,\ldots,x_n^q),(y_1,\ldots,y_n)\rangle=\sum_{i=1}^nx_i^qy_i$. 
Now, denote by $\mathcal{H}$ the associated matrix of $H$ with respect to the ordered $\F_{q^2}$-basis $(\alpha_1,\ldots,\alpha_n)$ and denote by $\mathcal{H}(i,j)$ its $(i,j)$-entry, namely, $\mathcal{H}(i,j)=H(\alpha_i,\alpha_j)=\mbox{Tr}(\alpha_j^q L(\alpha_i))$.

In the following we show how the codewords of the additive $d$-code in Theorem \ref{Th-Schmidt-5} can be expressed in the Hermitian matrix form. Note that
\begin{align*}
    H(x,y)&=\text{Tr}\left(\left(\sum_{i}y_i\alpha_i\right)^q\sum_{j}x_jL(\alpha_j)\right)
    =\mbox{Tr}\left( \sum_{i,j}y_i^qx_j\alpha_i^qL(\alpha_j)\right)\\
    &=\sum_{i,j}y_i^qx_j\mbox{Tr}\left(\alpha_i^qL(\alpha_j)\right)
    = \sum_{i,j}y_i^qx_j\mathcal{H}(i,j)
    =(y_1,\ldots,y_n)^q\cdot \mathcal{H}\cdot \left(\begin{array}{c}
         x_1  \\
         \vdots\\
         x_n
    \end{array}\right).
\end{align*}
In the literature, no encoding method has been given for Hermitian $d$-codes. In the following we show that the evaluation of the corresponding linearized polynomial at linearly independent elements $\alpha_1,\ldots,\alpha_n$ is a proper encoding method.

\smallskip 

Let $h=(h_1,\ldots,h_n)$ be the Hermitian vector corresponding to the Hermitian matrix $\mathcal{H}$, 
that is \[h_r=\sum_i \alpha_i H(\alpha_i,\alpha_r).\]
Since $H(\alpha_i,\alpha_r)=\mathrm{Tr}(\alpha_i^qf(\alpha_j))$, we can write $h_r$ as 
\[
    h_r=\sum_{i}\alpha_i \mbox{Tr}(\alpha_i^qL(\alpha_r))
    =\sum_{i}\alpha_i\mbox{Tr}\left(\alpha_i^q\sum_{j}c_j\alpha_j\right) =\sum_{i}\alpha_i\sum_{j}c_j\mbox{Tr}(\alpha_i^q\alpha_j)
    =\sum_{i}c_i\alpha_i,
\]
where $L(\alpha_r)=\sum_jc_j\alpha_j$,
by using the fact that $L(x)$ is linear over $\F_{q^2}$ and the fact that $(\alpha_1,\ldots,\alpha_n)$ is a Hermitian self-dual basis. From the above calculations we see that the evaluation encoding is the right way of encoding Hermitian $d$-codes, since $h_r=L(\alpha_r)$. 

\smallskip

Let $m=(n+1)/2$, $\kappa=(n-d)/2$ and $H$ be the Hermitian form given in Theorem \ref{Th-Schmidt-5}. Then the linearized polynomial in \eqref{ex:E}
can be written as 
$$L(x) = (b_0x)^{[m]} + \sum_{j=1}^{\kappa} \left( (b_jx)^{[m+j]}+b_j^{q} x^{[m-j]} \right).$$ 
Let $\{1,\eta\}$ be an $\F_{q^n}$-basis of $\F_{q^{2n}}$.
Let $\alpha_0,\alpha_1, \ldots, \alpha_{n-1}$ be a Hermitian self-dual basis of  $\F_{q^{2n}}$.
For the maximum Hermitian $d$-code in Theorem \ref{Th-Schmidt-5} and the evaluation points $\alpha_0,\alpha_1, \ldots, \alpha_{n-1}$, the encoding of
	a message $f=(f_0,\ldots, f_{k-1}) \in \F_{q^n}^k$ can be expressed as the evaluation of the following linearized polynomial at points $\alpha_0,\alpha_1, \ldots, \alpha_{n-1}$:
	\begin{equation}\label{eq-tilde-f}
	\begin{array}{rcl}
	    L(x) 
	    &=& (f_0x)^{[m]} + 
	     \left( \sum\limits_{j=1}^{\kappa} (f_j+\eta f_{\frac{k-1}2+j})^qx^{[m-j]} + 
	    (f_j+\eta f_{\frac{k-1}2+j}x)^{[m+j]}\right)
	    \\&=& \tilde{f}_0x + \tilde{f}_1x^{[1]} + \cdots + \tilde{f}_{n-1}x^{[n-1]}.
  \end{array}
	\end{equation}

Let $	M=
	\begin{pmatrix}
		\alpha_i^{[j]}
	\end{pmatrix}_{n\times n}$
	be the $n\times n$ \textit{Moore matrix} generated by $\alpha_i$'s. 

 So the encoding of the maximum Hermitian rank metric code can be expressed as \begin{equation}\label{eq1}
   (f_0,\ldots,f_{k-1})\mapsto (L(\alpha_0),\ldots,L(\alpha_{n-1}))=\tilde{f}\cdot M^T,
\end{equation}
where $\tilde{f}=(\tilde{f}_0, \ldots, \tilde{f}_{n-1})$ and $M^T$ is the transpose of the matrix $M$. 
As shown in \eqref{eq-tilde-f}, the first $m-\kappa$  and the last $ m-\kappa-2$ elements of $\tilde{f}$ are zero, we  only employ $k$ columns of the Moore matrix in the encoding process. 

\subsection{Interpolation Decoding}
	For a received word $r=c+e$ with an error $e$ added to the codeword $c$ during transmission, 
	when the error $e$ has rank $t\leq \lfloor \frac{d-1}{2}\rfloor$,
	the unique decoding task is to recover the unique codeword $c$ such that $d_r(c,r)\leq \lfloor \frac{d-1}{2}\rfloor$.
	
		\smallskip
	
	Suppose $g(x)=\sum_{i=0}^{n-1}g_ix^{[i]}$ is an error interpolation polynomial such that
	\begin{equation}\label{EqInterpolation}
	g(\alpha_i)=e_i=r_i-c_i, \quad i=0, \ldots, n-1.
	\end{equation}
	It is clear that the error vector  $e$ is uniquely determined by the polynomial $g(x)$, and denote $\tilde{g}=(g_0,\ldots, g_{n-1})$. From
	\eqref{eq1} and \eqref{EqInterpolation} it follows that
	$$
	r = c+e = (\tilde{f}+\tilde{g}) M^T.
	$$
	This is equivalent to
	\begin{align*}
	  r \cdot (M^T)^{-1}=&(0,\ldots,0,\tilde{f}_{m-\kappa}, \ldots,\tilde{f}_{m+\kappa}, 0,\ldots,0) 
	  \\& + (g_0, \ldots,g_{m-\kappa-1}, g_{m-\kappa},\ldots, g_{m+\kappa},g_{m+\kappa+1}, \ldots, g_{n-1}).
		\end{align*}
		where $\tilde{f}_m = f_0^{[m]}$ and for $j=1,2, \cdots, \kappa$, 
		$\tilde{f}_{m-j} = (f_j+\eta f_{\kappa+j})^q  \text{ and } \tilde{f}_{m+j} = \tilde{f}_{m-j}^{q^{n+2j}}$.
	Letting $\beta=(\beta_0, \ldots, \beta_{n-1})=r \cdot (M^T)^{-1}$, we obtain 
\begin{equation}\label{Eq-interpolation1}
\begin{cases}
	     (g_0, \ldots,g_{m-\kappa-1})=(\beta_0, \ldots,\beta_{m-\kappa-1}) \\
	     (g_{m+\kappa+1}, \ldots, g_{n-1})=(\beta_{m+\kappa+1}, \ldots, \beta_{n-1})
\end{cases}
\end{equation} 
Equivalently, we have 
\begin{equation}\label{Eq-interpolation2}
	     g_{m+\kappa + j} = \beta_{m+\kappa + j} \text{ for } j = 1, 2, \cdots, d-1,
\end{equation} where the subscripts are taken modulo $n$.
In addition, we have 
\begin{equation}\label{Eq-interpolation3}
	     g_{m-\kappa + j}  = \beta_{m-\kappa + j} - \tilde{f}_{m-\kappa + j}  \text{ for } j = 1, \ldots, 2\kappa. 
\end{equation}
With the relation $\tilde{f}_{m+j} = \tilde{f}_{m-j}^{q^{n+2j}}$ for $1\leq j\leq \kappa$, it is 
sufficient to recover the coefficients $g_{m+\kappa+1},\ldots,g_{m}$.

\smallskip

Therefore, the task of correcting error $e$ is equivalent to 
	reconstructing $g(x)$ from the available information characterized in the above relations  \eqref{Eq-interpolation2} and then apply $g(x)$ to recover $\tilde{f}$. This reconstruction process heavily depends on the 
	property of the associated  Dickson matrix of $g(x)$ and 
	will be discussed in next subsection.

\subsection{Polynomial reconstruction}
	
	The Dickson matrix associated with $g(x)$ can be given by 
	\begin{equation}\label{Eq-DM-G-Simplified}
	G=\begin{pmatrix}g^{[j]}_{i-j~({\rm mod~}n)} 
	\end{pmatrix}_{n\times n}
	= \left(G_0 \,\, G_1 \,\, \ldots \,\, G_{n-1}\right),
	\end{equation} 
	where the indices $i, \,j$ run through $\{0, 1, \ldots, n-1\}$ and 
	$G_{j}$ is the $j$-th column of $G$.
	
	 According to the properties of the Dickson matrix, when $D$ has rank $t$, any $t\times t$ matrix formed by $t$ successive rows and columns in $G$ is nonsingular, see e.g. \cite{WU201379} and \cite{Csajbok19}.
Then $G_0$ can be expressed as a linear combination 
of $G_1, \ldots, G_t$, namely,
$
G_{0} = \lambda_1 G_1 +\lambda_2 G_2 + \cdots + \lambda_{t} G_{t},
$
where $\lambda_1,\ldots, \lambda_{t}$ are elements in $\mathbb{F}_{q^{2n}}$.
This yields the following recursive equations
\begin{equation}\label{Eq-Gsub}
g_{i} = \lambda_1 g^{[1]}_{i-1} + \lambda_2 g^{[2]}_{i-2} + \cdots + \lambda_t g^{[t]}_{i-t}, \quad 0\leq i < n,
\end{equation} where the subscripts in $g_i$'s are taken modulo $n$.
Recall that the elements $g_{0},\ldots, g_{m-\kappa-1}$ and $g_{m+\kappa+1}, \ldots, g_{n-1}$ are known from \eqref{Eq-interpolation1}. 
Hence, we obtain the following linear equations with known coefficients and  variables $\lambda_1, \ldots, \lambda_{t}$:
\begin{equation}\label{Eq-Gsub-4}
g_{i} = \lambda_1 g^{[1]}_{i-1} + \lambda_2 g^{[2]}_{i-2} + \cdots + \lambda_t g^{[t]}_{i-t}, \,\, m+\kappa+t+1 \leq i<n+m-\kappa ~(\mbox{mod } n).
\end{equation} 

	The above recurrence gives a generalized version of $q$-linearized shift register as described in \cite{Sidorenk}, 
	where $(\lambda_1, \ldots, \lambda_{t})$ is the connection vector of the shift register.
    It is  the \textit{key equation} for the decoding algorithm in this paper, by which we shall
	reconstruct $g(x)$ in two major steps: 
	\begin{itemize}
		\item[]\noindent\textbf{Step 1.} derive  $\lambda_1, \ldots, \lambda_{t}$ from 
		\eqref{Eq-interpolation1} and \eqref{Eq-Gsub-4};
		\item[]\noindent\textbf{Step 2.} use $\lambda_1, \ldots, \lambda_{t}$ to compute $g_{m-\kappa}, \ldots, g_{m+\kappa}$ from \eqref{Eq-Gsub}.
	\end{itemize}
	Step 1 is the critical step in the decoding process, and Step 2 is simply a recursive process that can be done in linear time in $\mathbb{F}_{q^{2n}}$.
	The following discussion shows how the procedure of  Step 1 works. 
	
	As discussed in the beginning of this section, for an error vector with $\mbox{rk}(e)=t \leq \lfloor \frac{d-1}{2}\rfloor$,  \eqref{Eq-Gsub-4} contains a system of
	$(n-k)-t=d-1-t\geq t$ affine linear equations 
	in the variables $\lambda_1, \ldots, \lambda_{t}$, which has rank $t$. Hence the variables $\lambda_1, \ldots, \lambda_{t}$ can be uniquely determined.	Here we assume the code has high code rate, for which the Berlekamp-Massey algorithm is more efficient. 
	Although the recurrence equation \eqref{Eq-Gsub-4} is a generalized version of the ones in \cite{Richter,Sidorenk}, the modified Berlekamp-Massey algorithm can be applied here to recover the 
	coefficients $\lambda_1, \ldots, \lambda_{t}$. 

	\smallskip

\section{Conclusion}

In this abstract we have proposed both interpolation-based encoding and decoding algorithms for a family of maximum Hermitian rank metric codes when the length and the minimum distance of the code are both odd. 
In the future we will extend it to other classes of maximum rank distance codes with restrictions.

\bibliographystyle{abbrv}
\bibliography{RankMetricCodes}

\end{document}